%% file: template.tex
\documentclass[a4paper]{article}

\usepackage[fussy]{INTERSPEECH2020}
\usepackage{hyperref}
\title{DurIAN-SC: Duration Informed Attention Network based Singing Voice Conversion System}
\name{Liqiang Zhang\sthanks{*Work performed while interning at Tencent AI Lab} $^1$, 
Chengzhu Yu$^2$, Heng Lu$^2$, Chao Weng$^2$, Chunlei Zhang$^2$, Yusong Wu$^2$, \\
\textit{Xiang Xie$^1$, Zijin Li$^3$, Dong Yu$^2$}}
\address{
  $^1$Beijing Institute of Technology\\
  $^2$Tencent AI Lab\\
  $^3$China Conservatory of Music}
\email{\{zhlq,xiexiang\}@bit.edu.cn,\{czyu,bearlu,cweng,cleizhang,ysw,dyu\}@tencent.com}

\begin{document}

\maketitle
\begin{abstract}
Singing voice conversion is converting the timbre in the source singing to the target speaker's voice while keeping singing content the same. However, singing data for target speaker is much more difficult to collect compared with normal speech data.
In this paper, we introduce a singing voice conversion algorithm that is capable of generating high quality target speaker's singing using only his/her normal speech data. First, we manage to integrate the training and conversion process of speech and singing into one framework by unifying the features used in standard speech synthesis system and singing synthesis system. In this way, normal speech data can also contribute to singing voice conversion training, making the singing voice conversion system more robust especially when the singing database is small.
Moreover, in order to achieve one-shot singing voice conversion, a speaker embedding module is developed using both speech and singing data, which provides target speaker identify information during conversion. 
Experiments indicate proposed sing conversion system can convert source singing to target speaker's high-quality singing with only 20 seconds of target speaker's enrollment speech data. 
\end{abstract}

\section{Introduction}

Singing is one of the predominant form of the music arts and singing voice conversion and synthesis can have many potential applications in entertainment industries. Over the past decades, many methods have been proposed to increase the naturalness of synthesized singing. These include the methods based on unit selection and concatenation\cite{bonada2016expressive} as well as more the recent approaches based on deep neural network (DNN) \cite{nishimura2016singing} and auto-regressive generation models \cite{blaauw2017neural}. 

While existing singing synthesis algorithms are able to producing natural singing, it basically requires large amount of singing data from one same speaker in order to generate his/her singing. Comparing to normal speech data collection, singing data is much more difficult and more expensive to obtain. To alleviate such limitations, data efficient singing synthesis approaches \cite{blaauw2019data} have been proposed recently. In \cite{blaauw2019data}, a large singing synthesis model trained from multi-speaker is adaptively fine-tuned with a small amount of target speaker's singing data to generate the target singing model. Alternatively, singing generation for new voices can be achieved through singing voice conversion. The goal of singing voice conversion is to convert the source singing to the timbre of target speaker while keeping singing content untouched. Traditional singing voice conversion  \cite{kobayashi2014statistical,kobayashi2015statistical,villavicencio2010applying} relies on parallel singing data to learn conversion functions between different speakers. However, a recent study \cite{nachmani2019unsupervised} proposed an unsupervised singing voice conversion method based on WaveNet \cite{oord2016wavenet} autoencoder architecture to achieve non parallel singing voice conversion. In \cite{nachmani2019unsupervised}, neither singing data nor the transcribed lyrics or notes is needed. 

While above mentioned methods could efficiently generate singing with new voices, they still require an essential amount of singing voice samples from target speakers. This limits the applications of singing generation to relatively restricted scenarios where there has to be target speaker's singing data. On the other hand, normal speech samples are much easier to collect than singing. There are only limited studies on investigating to use normal speech data to enhance singing generation. The speech-to-singing synthesis method proposed in \cite{saitou2007speech} attempts to convert a speaking voice to singing by directly modifying acoustic features such as \textit{f0} contour and phone duration extracted from reading speech. While speech-to-singing approaches could produce singing from reading lyrics, it normally requires non-trivial amount of manual tuning of acoustic features for achieving high intelligibility and naturalness of singing voices. 


\begin{figure*}[t]
  \centering
  \includegraphics[clip, trim=0.6cm 0.6cm 0.6cm 0.8cm, width=0.9\textwidth]{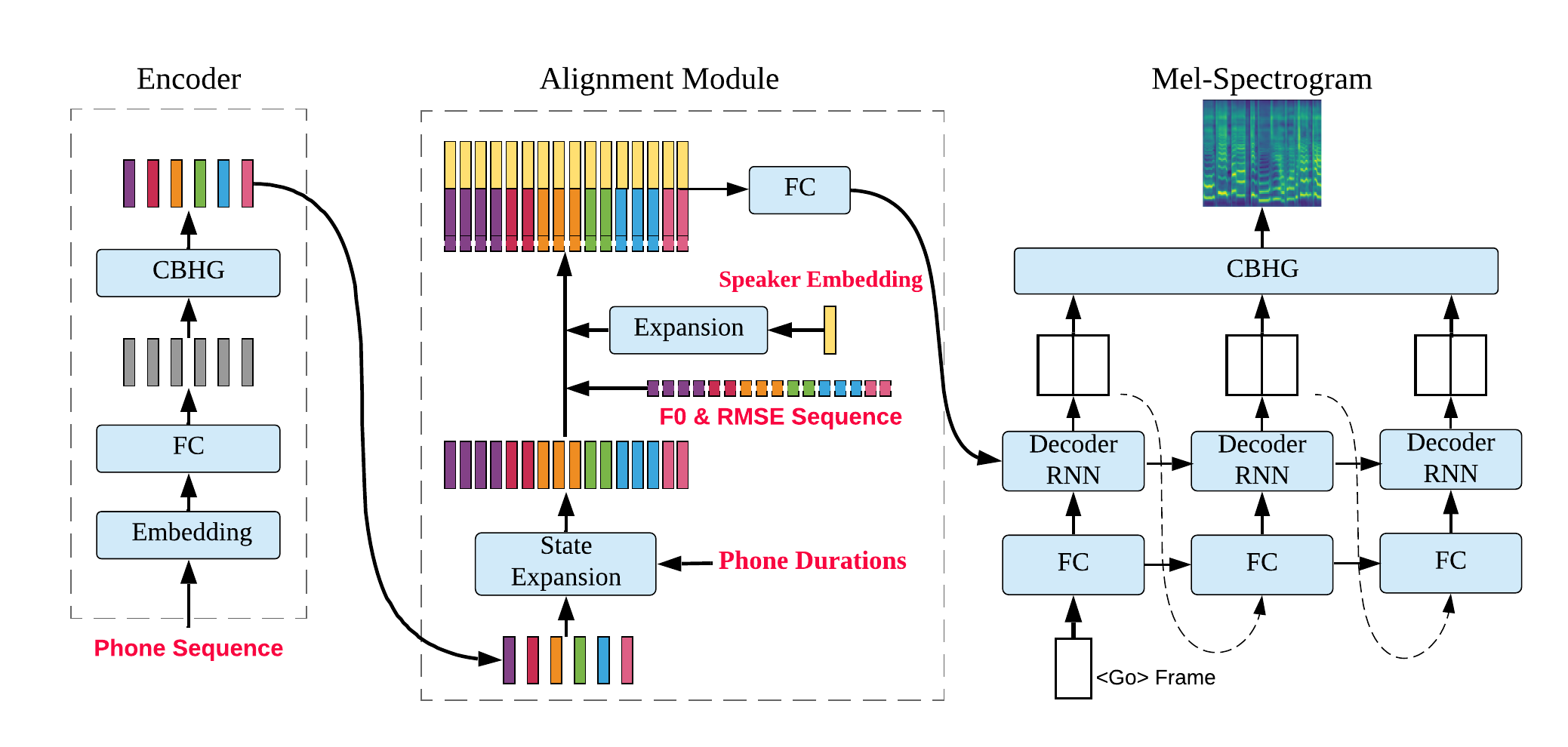}
  \caption{Model architecture of DurIAN-SC. RMSE means root mean square energy, FC represents the full connected layer, Expansion means expanding the time dimension to frame level.}
\vspace{-5mm}
\label{fig:f2}
\end{figure*}

Duration Informed Attention Network (DurIAN)\cite{yu2019durian}, originally proposed for the task of multimodal synthesis, is essentially an autoregressive feature generation framework that can generate acoustic features (e.g., mel-spectrogram) for any audio source frame by frame. In this paper, we proposed a DurIAN based speech and singing voice conversion system (DurIAN-SC), a unified speech and singing conversion framework\footnote{Sound demo of proposed algorithm can be found at \url{https://tencent-ailab.github.io/learning\_singing\_from\_speech}}. There are two major contributions for the proposed method: 1) Despite the input feature for conventional speech synthesis and singing synthesis is different, proposed framework unifies the training process for both speech and singing synthesis. Thus in this work, we can even train the singing voice conversion model just using speech data. 2) Instead of the commonly used trainable Look Up Table (LUT)\cite{nachmani2019unsupervised} for speaker embedding, we use a pre-trained speaker embedding network module for speaker d-vector\cite{Gonzalez2014Deep,snyder2018x} extraction. Extracted speaker d-vectors are then fed into singing voice conversion network 
as the speaker embedding to represent the speaker identity. During conversion, only 20 seconds speech or singing data is needed for the tester's d-vector extraction.  Experiments show proposed algorithm can generates high-quality singing voices when using only speech data. The Mean Opinion Scores (MOS) of naturalness and similarity indicates our system can perform one-shot singing voice conversion with only 20 seconds tester's speech data.

The paper is organized as following. Section 2 introduces the architecture of our proposed  conversion model. Experiments are introduced in Section 3. And section 4 is the conclusion.

\section{Model Architecture}


\subsection{DurIAN-SC}
\label{ssec:subhead}
While DurIAN was originally proposed for the task of multimodal speech synthesis, it has many advantages over conventional End-to-End framework, especially for its stable in synthesis and its duration controllability. The original DurIAN model is modified here to perform speech and singing synthesis at the same time. Here we use text/song lyric as one of input for both speech and singing data. Text or song lyric is then transferred to phone sequence with prosody token by text-to-speech TTS front-end module. The commonly used music score is not used in our singing voice conversion framework. Instead, we use frame level \textit{f0} and average Root Mean Square Energy (RMSE) extracted from both original singing/speech as additional input conditions (Fig.~\ref{fig:f2}). For singing voice conversion, the \textit{f0} and rhythm is totally decided by score notes and the content itself, and this is the part we do not convert unless there is large gap between source and target speaker's singing pitch range. Further, we found that if using RMSE as input condition in training, the loss convergence would be much faster.

The architecture of DurIAN-SC is illustrated in Fig.~\ref{fig:f2}. It includes (1) an encoder that encodes the context of each phone, (2) an alignment model that aligns the input phone sequence and to target acoustic frames, (3) an auto-regressive decoder network that generates target mel-spectrogram features frame by frame. 

\subsubsection{Encoder}
We use phone sequence $\mathrm{x_{1:N}}$ directly as input for both speech and singing synthesis. The output of the encoder  $\mathrm{h_{1:N}}$ is a sequence of hidden states containing the sequential representation of the input phones as
\begin{equation}
\mathrm{h_{1:N}} = \mathrm{encoder}(\mathrm{x_{1:N}})
\end{equation}
where $\mathrm{N}$ is the length of input phone sequences, encoder module contains a phone embedding, fully connected layers and a CBHG\cite{wang2017tacotron} module, which is a combination module of Convolution layer, Highway network\cite{srivastava2015highway} and bidirectional GRU\cite{chung2014empirical}.

\subsubsection{Alignment model} 
The purpose of alignment model is to generate frame aligned hidden states which is further fed into auto-regressive decoder. Here, the output hidden sequence from encoder $\mathrm{h_{1:N}}$ is first expanded according to the duration of each phone as  
\begin{equation}
\mathrm{e_{1:T}} = \mathrm{state\_expand}(\mathrm{h_{1:N}}, \mathrm{d_{1:N}})
\end{equation}
where $T$ is the total number of input audio frames. The state expansion is simply the replication of hidden states according to the provided phone duration $\mathrm{d_{1:N}}$. The duration of each phone is obtained from force alignments performed on input source phones and acoustic features sequences. The frame aligned hidden states $\mathrm{e_{1:T}}$ is then concatenated with frame level \textit{f0}, RMSE and speaker embedding, as we can see in Fig.~\ref{fig:f2}.
\begin{equation}\label{eq5}
\mathrm{e_{1:T}^{'}=FC(e_{1:T} \vee f_{1:T} \vee r_{1:T} \vee D_{1:T} )} 
\end{equation}
where $\vee$ indicates concatenation,$\mathrm{FC}$ indicates the fully connected layer, $\mathrm{f_{1:T}}$ represents \textit{f0} for each frame, $\mathrm{D_{1:T}}$ represents the speaker embedding expanded to frame level. And $\mathrm{r_{1:T}}$ is the RMSE for each frame.

\subsubsection{Decoder} 
The decoder is the same as in DurIAN, composed of two auto-regressive RNN layers. Different from the attention mechanism used in the end-to-end systems, the attention context here is computed from a small number of encoded hidden states that are aligned with the target frames, which reduces the artifacts observed in the end-to-end system\cite{wang2017tacotron}. We decode two frames per time step in our system. The output from the decoder network $\mathrm{y_{1:T}^{'}}$ is passed through a post-CBHG \cite{wang2017tacotron} to improve the quality of predicted mel-spectrogram as
\begin{equation}\label{eq6}
\mathrm{y_{1:T}^{'}=decoder(e_{1:T}^{'})}
\end{equation}
\begin{equation}\label{eq7}
\mathrm{\hat{y}_{1:T}=cbhg(y_{1:T}^{'})}
\end{equation}
The entire network is trained to minimize the mel-spectrogram prediction loss the same as in DurIAN.



\subsection{Singing Voice Conversion Process}
\begin{figure}[t]
  \centering
  \includegraphics[clip,trim=0.2cm 0.8cm 0.2cm 0.2cm, width=0.5\textwidth]{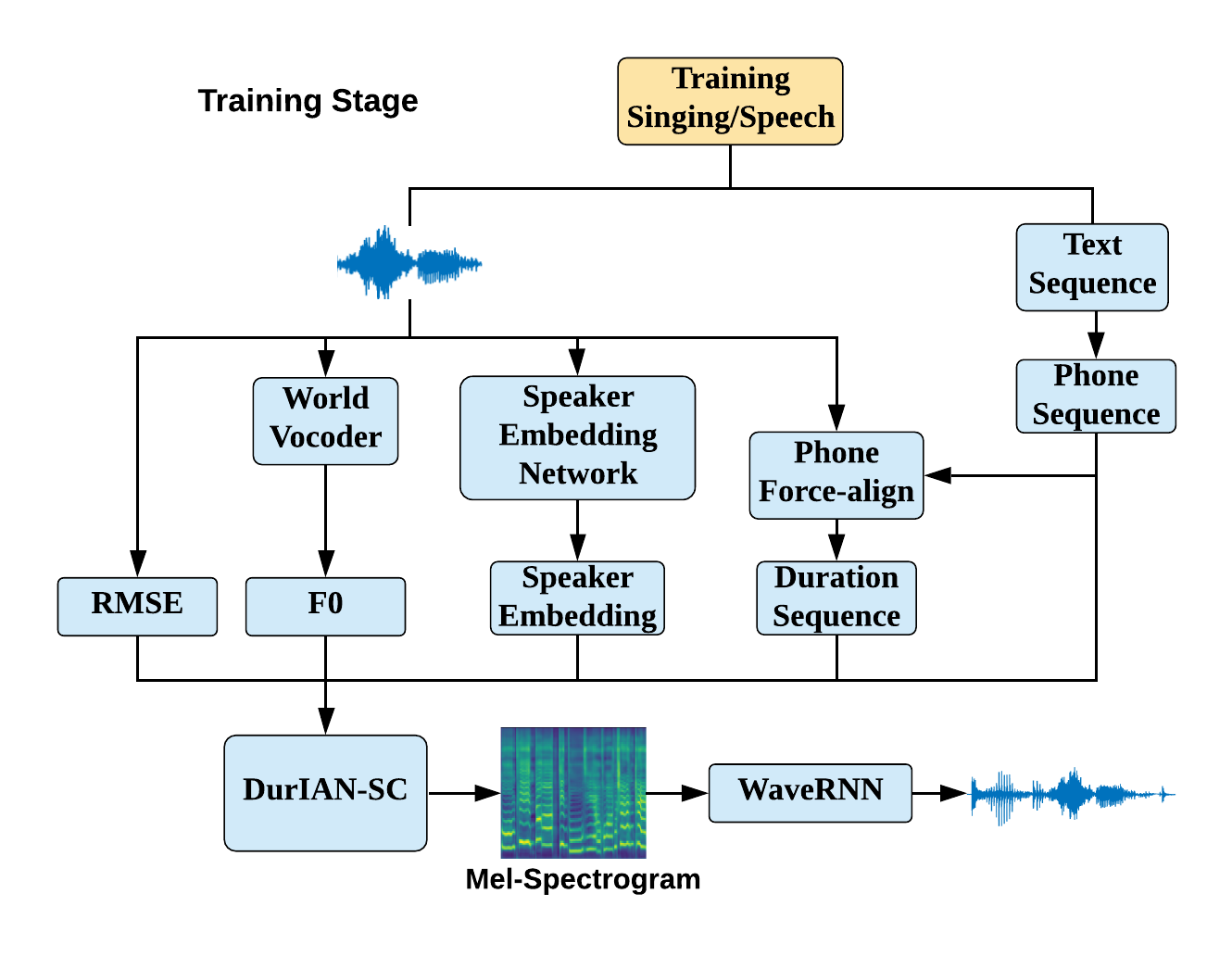}
\caption{The process diagram of training stage. The WaveRNN \cite{DBLP:journals/corr/abs-1802-08435} model is trained separately.}
\vspace{-5mm}
\label{fig:f1}
\end{figure}

\begin{figure}[t]
  \centering
  \includegraphics[clip, trim=0.2cm 0.65cm 0.2cm 0.2cm, width=0.5\textwidth]{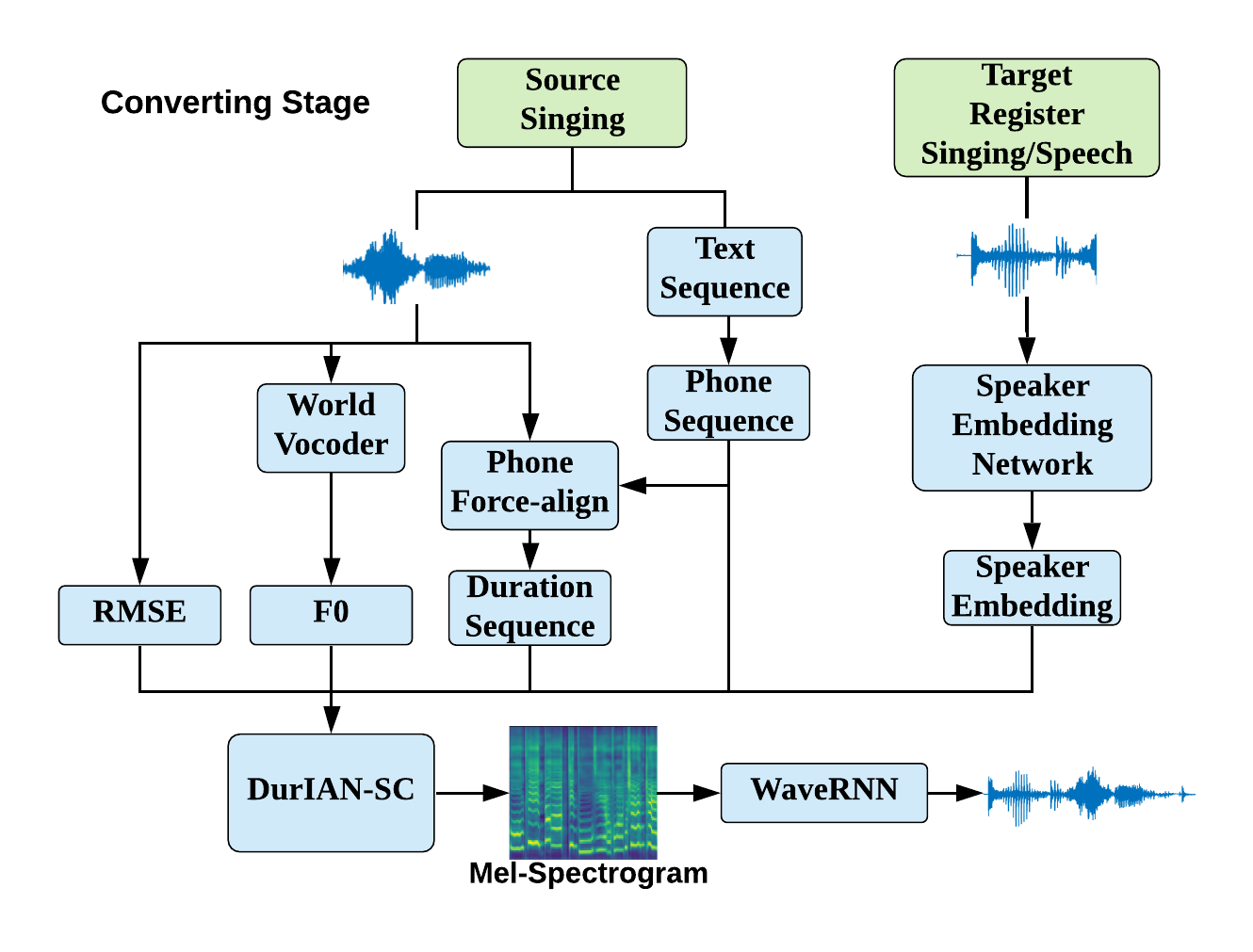}
\caption{The process diagram of converting stage.}
\vspace{-6mm}
\label{fig:f3}
\end{figure}

The training stage is illustrated in Fig.~\ref{fig:f1}, and the converting stage is illustrated in Fig.~\ref{fig:f3}.

\subsubsection{Data Preparation}
Our training dataset is composed a mix of normal speech data and singing data. TTS front-end is used to parse text or song lyrics into phone sequence. Acoustic feature including mel-sepctrogram, \textit{f0} and RMSE are extracted for every frame of training data. Note that the \textit{f0} is extracted with World vocoder\cite{morise2016world}. Since DurIAN structure needs phone alignment as input, a Time delay neural network (TDNN) is employed here to 
force-align the extracted acoustic feature with phone sequence. Different from normal TTS for Mandarin which use phone identity plus 5 tones in the modeling, non-tonal phones are used in our experiment to bridge the gap between speech phones and singing phones. Finally, phone duration can be extracted from the aligned phone sequence.

\subsubsection{Speaker embedding network}
To provide the DurIAN-SC with robust speaker embedding on Mandarin language. External Mandarin corpora are explored to train a speaker embedding network, which is then used as a pre-trained module. The external training set contains of 8800 speaker drawn from two gender-balanced public speech recognition datasets\footnote{http://en.speechocean.com/datacenter/details/254.htm}. The training data is then augmented 2 folds to incorporate variabilities from distance (reverberation),  channel or background noise, resulting in a training pool with 2.8M utterances. 257-d raw short time fourier transform (STFT) features are extracted with a 32ms window and the time shift of feature frames is 16ms. The non-speech part is removed by a energy based voice activity detection. The utterance is randomly segmented into 100-200 frames to control the duration variability in the training phase. For the choice of network architecture, we employ a TDNN framework which is similar to \cite{snyder2018x,ji2020}. The speaker embedding training guilded with a multi-task loss, which employs both the large margin cosine loss (LMCL) and the triplet loss \cite{zhang2017end,wang2018cosface,zhang2019utd}. 

In order to further boost the capability for singing data, the internal singing corpus is incorporated in the speaker embedding training. Since the singing corpus is not provided with speaker label, we employ a bottom-up hierarchical agglomerative clustering (HAC) to assign a pseudo speaker label for each singing segment. Specifically, we first extract speaker embedding for singing corpus using the external speaker embedding model. Then, HAC is applied to produce 1000 speaker ``IDs" from the training singing corpus (3500 singing segments). Finally, the clustered corpus is pooled with external speech data for another round of speaker embedding training. The final system is utilized to extract speaker embedding for speech/singing.      

\subsubsection{Training and conversion process}
In the training stage, both the normal speech and singing data could be used as input training data. The \textit{f0}, RMSE, phone sequence and phone duration are extracted as shown in section 2.2.1. Speaker embedding are extracted using the pre-trained speaker embedding network introduced in the previous section. DurIAN-SC model is then trained based on these extracted acoustic features and speaker embedding. 

In singing voice conversion stage, \textit{f0}, RMSE and phone duration are extracted from source singing and later used in conversion process as condition. Using the pre-trained speaker embedding network, target speaker embedding can be obtained by testing on target speaker's singing or speech data with a length of only 20 seconds. By conditioning on the extracted target speaker embedding, mel-spectrogram can be generated with target speaker's timbre through the model trained in the last session. Finally, WaveRNN \cite{DBLP:journals/corr/abs-1802-08435} is employed as Neural Vocoder for waveform generation.   

In case there is large gap between source and target speaker's singing pitch range, which often happen when performing cross gender conversion, we shift original source key linearly to make it easier for target speaker to 'sing' the same song as source. The input \textit{f0} is multiplied by a factor $\nu$ as:

\begin{equation}\label{eq1}
\nu  =  \frac{\mathit{mean}(x^{t})}{\mathit{mean}(x^{s})}
\end{equation}
where $x^{s}$ is the source singing \textit{f0}, $x_{i}^{t}$ is the target register speech or singing \textit{f0}. $\mathit{mean}$ represents to average  \textit{f0} across all vowel phones in all the audios by the source or target speaker.

\section{Experiments}


\subsection{Dataset}

Two databases are used in our experiments. Database A is a  large multi-singer Mandarin singing corpus containing 18-hour singing data. There are 3600 singing segments from various songs in corpus A, and each with an average length of 20 seconds. Each singing fragment is by a different singer. Amongst all singing fragments, 2600 are by female singers and 1000 are by male singers. This multi-singer singing corpus are recorded by singers themselves with various recording devices. All songs are down sampled to 24kHz. 

Database B is speech database containing 10-hour multi-speaker Mandarin normal TTS speech data. There are 3 male speakers and 4 female speakers in this corpus, each with a duration around 1.5 hours. The sampling rate is also set to 24kHz.

In the singing voice conversion experiments, all source singing is chosen randomly from another mandarin singing corpus C.

\subsection{Model Hyperparameters}

In our experiment, the dimensions of the phone embedding, speaker embedding, encoder CBHG module, attention layer are all set to 256. The decoder has 2 GRU layers with 256 dimension and batch normalization is used in the encoder and post-net module. We use Adam optimizer and $0.001$ initial learning rate with warm-up \cite{goyal2017accurate} schedule. In training stage, a total of 250k steps with a batch size of 32 were trained till convergence.

\subsection{Naturalness and Similarity Evaluation}

In the singing voice conversion test, Mean Opinion Scores (MOS) on naturalness and similarity to target speaker are evaluated. The scale of MOS is set between 1 to 5 with 5 representing the best performance and 1 the worst. 10 testers participated in our listening test.

\subsubsection{Experiment on speaker embedding representation } 
In this experiment, we compare the singing naturalness and similarity to target speaker by proposed d-vector based speaker embedding and LUT based trainable speaker embedding. Two systems are built respectively. The training dataset used here is the 18-hour singing database A introduced in section 3.1. We use a total of 3500 singing fragments in training. In testing, 3 female and 3 male singers are randomly chosen from training set for in-set test. To evaluate the out-set singing voice conversion performance, 4 speakers from the speech dataset B are chosen for test. Here, only a 20s period of singing or speech data are used from each testers for speaker d-vector extraction. As the baseline system, the LUT based trainable speaker embedding is trained alongside the singing voice conversion DurIAN-SC model. The out-of-set baseline system is not tested because baseline system can not convert to unseen target. 

\begin{table}[htb!]
  \caption{Comparison of speaker embedding extraction methods: LUT and speaker D-vector. The 'Target Singer' column indicates whether target speaker's singing data is used in training.}
  \label{tab:ex1}
  \centering
  \begin{tabular}{cccc}
    \toprule
    \textbf{Method} & \textbf{Target Singer} & \textbf{Naturalness} & \textbf{Similarity}                \\
    \midrule
            D-vector &in-set & 3.70 & 3.61 \\
			LUT &in-set & 3.61 & 3.56 \\
			D-vector &out-of-set & 3.69 & 3.10 \\
			LUT &out-of-set & - & - \\
    \bottomrule
  \vspace{-8mm}
  \end{tabular}
\end{table}

As shown in Table \ref{tab:ex1}, for the in-set test, proposed D-vector speaker embedding system outperforms the baseline LUT speaker embedding system in both MOS naturalness and similarity by a small margin. The result is in line with expectations. For the baseline trainable LUT speaker embedding system, the speaker embedding is trained alongside the singing voice conversion model, that makes the total free parameter in the system is actually more than proposed method especially for the 'seen' speaker. However on the other side, because there is only 20 seconds data per each singer in the training, it could be hard for the trainable LUT speaker embedding method to learn a really good speaker embedding. Meanwhile, proposed speaker embedding network is an independent module which is pre-trained on a lot extra speaker recognition data. While for the out-set test, the MOS scores for proposed method is lower than in-set test especially on similarity. We believe this is normal result for the model parameters are not fine-tuned with the 'unseen' speaker's data. And speaker d-vectors are extracted from only 20 seconds of target speaker's register speech or singing. At least, unlike the baseline system, proposed method save the trouble to fine-tune and update model parameters for each new user. 


\subsubsection{Using speech corpus in singing voice conversion} 

To demonstrate proposed system can learn singing voice conversion from only speech data, three different systems are trained using: 1) only speech data, 2) mix of speech and singing data, and 3) singing data only, respectively for comparison.

\begin{table}[htb!]
  \caption{Singing voice conversion experiments trained with speech data. Dataset indicates the type of training data.}
  \label{tab:ex2}
  \centering
  \begin{tabular}{cccc}
    \toprule
    \textbf{Dataset}  &  \textbf{Naturalness} & \textbf{Similarity}               \\
    \midrule
            Speech $\&$ Singing & 3.71 & 3.74 \\
			Only Speech  & 3.65 & 3.71 \\
			Only Singing  & 3.70 & 3.61 \\
    \bottomrule
  \vspace{-5mm}
  \end{tabular}
\end{table}

Results in Table \ref{tab:ex2} show that all three above mentioned systems has close performance. This interesting result indicates that in the proposed system, speech data can contribute equally to singing voice conversion as singing data. In this case, we can use only speech data when target's singing data is not available. In our experiments, it is noticed that by adding some speech data to singing voice conversion training process, the generated target singing will have clearer pronunciation. Speech data in training also helps to improve the singing voice conversion similarity.

\section{Conclusion}

In this paper, we proposed an singing voice conversion model DurIAN-SC with a unified framework of speech and singing data. For those speakers with none singing data, our method could convert to their singings by training on only their speech data. Through a pre-trained speaker embedding network, we could convert to 'unseen' speakers' singing with only a 20 second length of data. Experiments indicate the proposed model can generate high-quality singing voices for in-set 'seen' target speakers in terms of both naturalness and similarity. In the meanwhile, proposed system can also one-shot convert to out-of-set 'unseen' users with small register data. In the future work, we will continue to make our model nore robust and improve the similarity of the 'unseen' singing voice conversion.

\input{template.bbl}

\bibliographystyle{IEEEtran}


\end{document}

%% file: template.bbl

%% file: template.bbl
\begin{thebibliography}{10}
\providecommand{\url}[1]{#1}
\csname url@samestyle\endcsname
\providecommand{\newblock}{\relax}
\providecommand{\bibinfo}[2]{#2}
\providecommand{\BIBentrySTDinterwordspacing}{\spaceskip=0pt\relax}
\providecommand{\BIBentryALTinterwordstretchfactor}{4}
\providecommand{\BIBentryALTinterwordspacing}{\spaceskip=\fontdimen2\font plus
\BIBentryALTinterwordstretchfactor\fontdimen3\font minus
  \fontdimen4\font\relax}
\providecommand{\BIBforeignlanguage}[2]{{%
\expandafter\ifx\csname l@#1\endcsname\relax
\typeout{** WARNING: IEEEtran.bst: No hyphenation pattern has been}%
\typeout{** loaded for the language `#1'. Using the pattern for}%
\typeout{** the default language instead.}%
\else
\language=\csname l@#1\endcsname
\fi
#2}}
\providecommand{\BIBdecl}{\relax}
\BIBdecl

\bibitem{bonada2016expressive}
J.~Bonada, M.~Umbert, and M.~Blaauw, ``Expressive singing synthesis based on
  unit selection for the singing synthesis challenge 2016.'' in
  \emph{INTERSPEECH}, 2016, pp. 1230--1234.

\bibitem{nishimura2016singing}
M.~Nishimura, K.~Hashimoto, K.~Oura, Y.~Nankaku, and K.~Tokuda, ``Singing voice
  synthesis based on deep neural networks.'' in \emph{Interspeech}, 2016, pp.
  2478--2482.

\bibitem{blaauw2017neural}
M.~Blaauw and J.~Bonada, ``A neural parametric singing synthesizer modeling
  timbre and expression from natural songs,'' \emph{Applied Sciences}, vol.~7,
  no.~12, p. 1313, 2017.

\bibitem{blaauw2019data}
M.~Blaauw, J.~Bonada, and R.~Daido, ``Data efficient voice cloning for neural
  singing synthesis,'' in \emph{ICASSP 2019-2019 IEEE International Conference
  on Acoustics, Speech and Signal Processing (ICASSP)}.\hskip 1em plus 0.5em
  minus 0.4em\relax IEEE, 2019, pp. 6840--6844.

\bibitem{kobayashi2014statistical}
K.~Kobayashi, T.~Toda, G.~Neubig, S.~Sakti, and S.~Nakamura, ``Statistical
  singing voice conversion with direct waveform modification based on the
  spectrum differential,'' in \emph{Fifteenth Annual Conference of the
  International Speech Communication Association}, 2014.

\bibitem{kobayashi2015statistical}
------, ``Statistical singing voice conversion based on direct waveform
  modification with global variance,'' in \emph{Sixteenth Annual Conference of
  the International Speech Communication Association}, 2015.

\bibitem{villavicencio2010applying}
F.~Villavicencio and J.~Bonada, ``Applying voice conversion to concatenative
  singing-voice synthesis,'' in \emph{Eleventh Annual Conference of the
  International Speech Communication Association}, 2010.

\bibitem{nachmani2019unsupervised}
E.~Nachmani and L.~Wolf, ``Unsupervised singing voice conversion,'' \emph{arXiv
  preprint arXiv:1904.06590}, 2019.

\bibitem{oord2016wavenet}
A.~v.~d. Oord, S.~Dieleman, H.~Zen, K.~Simonyan, O.~Vinyals, A.~Graves,
  N.~Kalchbrenner, A.~Senior, and K.~Kavukcuoglu, ``Wavenet: A generative model
  for raw audio,'' \emph{arXiv preprint arXiv:1609.03499}, 2016.

\bibitem{saitou2007speech}
T.~Saitou, M.~Goto, M.~Unoki, and M.~Akagi, ``Speech-to-singing synthesis:
  Converting speaking voices to singing voices by controlling acoustic features
  unique to singing voices,'' in \emph{2007 IEEE Workshop on Applications of
  Signal Processing to Audio and Acoustics}.\hskip 1em plus 0.5em minus
  0.4em\relax IEEE, 2007, pp. 215--218.

\bibitem{yu2019durian}
C.~Yu, H.~Lu, N.~Hu, M.~Yu, C.~Weng, K.~Xu, P.~Liu, D.~Tuo, S.~Kang, G.~Lei
  \emph{et~al.}, ``Durian: Duration informed attention network for multimodal
  synthesis,'' \emph{arXiv preprint arXiv:1909.01700}, 2019.

\bibitem{Gonzalez2014Deep}
J.~Gonzalez-Dominguez, ``Deep neural networks for small footprint
  text-dependent speaker verification,'' in \emph{ICASSP 2014 - 2014 IEEE
  International Conference on Acoustics, Speech and Signal Processing
  (ICASSP)}, 2014.

\bibitem{snyder2018x}
D.~Snyder, D.~Garcia-Romero, G.~Sell, D.~Povey, and S.~Khudanpur, ``X-vectors:
  Robust dnn embeddings for speaker recognition,'' in \emph{2018 IEEE
  International Conference on Acoustics, Speech and Signal Processing
  (ICASSP)}.\hskip 1em plus 0.5em minus 0.4em\relax IEEE, 2018, pp. 5329--5333.

\bibitem{wang2017tacotron}
Y.~Wang, R.~Skerry-Ryan, D.~Stanton, Y.~Wu, R.~J. Weiss, N.~Jaitly, Z.~Yang,
  Y.~Xiao, Z.~Chen, S.~Bengio \emph{et~al.}, ``Tacotron: Towards end-to-end
  speech synthesis,'' \emph{arXiv preprint arXiv:1703.10135}, 2017.

\bibitem{srivastava2015highway}
R.~K. Srivastava, K.~Greff, and J.~Schmidhuber, ``Highway networks,'' 2015.

\bibitem{chung2014empirical}
J.~Chung, C.~Gulcehre, K.~Cho, and Y.~Bengio, ``Empirical evaluation of gated
  recurrent neural networks on sequence modeling,'' \emph{arXiv preprint
  arXiv:1412.3555}, 2014.

\bibitem{DBLP:journals/corr/abs-1802-08435}
\BIBentryALTinterwordspacing
N.~Kalchbrenner, E.~Elsen, K.~Simonyan, S.~Noury, N.~Casagrande, E.~Lockhart,
  F.~Stimberg, A.~van~den Oord, S.~Dieleman, and K.~Kavukcuoglu, ``Efficient
  neural audio synthesis,'' \emph{CoRR}, vol. abs/1802.08435, 2018. [Online].
  Available: \url{http://arxiv.org/abs/1802.08435}
\BIBentrySTDinterwordspacing

\bibitem{morise2016world}
M.~Morise, F.~Yokomori, and K.~Ozawa, ``World: a vocoder-based high-quality
  speech synthesis system for real-time applications,'' \emph{IEICE
  TRANSACTIONS on Information and Systems}, vol.~99, no.~7, pp. 1877--1884,
  2016.

\bibitem{ji2020}
X.~{Ji}, M.~{Yu}, C.~{Zhang}, D.~{Su}, T.~{Yu}, X.~{Liu}, and D.~{Yu},
  ``Speaker-aware target speaker enhancement by jointly learning with speaker
  embedding extraction,'' in \emph{ICASSP 2020 - 2020 IEEE International
  Conference on Acoustics, Speech and Signal Processing (ICASSP)}, 2020, pp.
  7294--7298.

\bibitem{zhang2017end}
C.~Zhang and K.~Koishida, ``End-to-end text-independent speaker verification
  with triplet loss on short utterances,'' in \emph{Proc. Interspeech 2017},
  2017, pp. 1487--1491.

\bibitem{wang2018cosface}
H.~Wang, Y.~Wang, Z.~Zhou, X.~Ji, D.~Gong, J.~Zhou, Z.~Li, and W.~Liu,
  ``Cosface: Large margin cosine loss for deep face recognition,'' in
  \emph{Proceedings of CVPR}, 2018, pp. 5265--5274.

\bibitem{zhang2019utd}
C.~Zhang, F.~Bahmaninezhad, S.~Ranjan, H.~Dubey, W.~Xia, and J.~H. Hansen,
  ``Utd-crss systems for 2018 nist speaker recognition evaluation,'' in
  \emph{ICASSP 2019-2019 IEEE International Conference on Acoustics, Speech and
  Signal Processing (ICASSP)}.\hskip 1em plus 0.5em minus 0.4em\relax IEEE,
  2019, pp. 5776--5780.

\bibitem{goyal2017accurate}
P.~Goyal, P.~Doll{\'a}r, R.~Girshick, P.~Noordhuis, L.~Wesolowski, A.~Kyrola,
  A.~Tulloch, Y.~Jia, and K.~He, ``Accurate, large minibatch sgd: Training
  imagenet in 1 hour,'' \emph{arXiv preprint arXiv:1706.02677}, 2017.

\end{thebibliography}
